\documentclass[12pt,preprint]{aastex}

\slugcomment{accepted by ApJL, May 11, 2005}

\begin{document}

\title{Towards Planetesimals in the Disk around TW~Hya: \\
$\lambda=3.5$~centimeter Dust Emission}

\author{D.J. Wilner\altaffilmark{1}
        P. D'Alessio\altaffilmark{2}
        N. Calvet\altaffilmark{1}
        M. J. Claussen\altaffilmark{3}
        L. Hartmann\altaffilmark{1}
       }
\email{dwilner@cfa.harvard.edu}
\altaffiltext{1}{Harvard-Smithsonian Center for Astrophysics,
  60 Garden Street, Cambridge, MA 02138, USA}
\altaffiltext{2}{Instituto de Astronomia, UNAM, Apartado Postal 72-3, 
  58089 Morelia, Michoacan, Mexico}
\altaffiltext{3}{National Radio Astronomy Observatory, P.O. Box O,  
  Socorro, NM 87801}

\begin{abstract}
We present Very Large Array observations at $\lambda=3.5$ cm of the
nearby young star TW~Hya that show the emission is constant in time 
over weeks, months and years, and spatially resolved with peak 
brightness temperature $\sim10$~K at $\sim0\farcs25$ (15 AU) resolution. 
These features are naturally explained if the emission mechanism
at this wavelength is thermal emission from dust particles in the disk 
surrounding the star.
To account quantitatively for the observations, we construct a 
self-consistent accretion disk model that incorporates a population 
of centimeter size particles that matches the long wavelength spectrum 
and spatial distribution. A substantial mass fraction of orbiting 
particles in the TW~Hya disk must have agglomerated to centimeter size. 
These data provide the first clear indication that dust emission from 
protoplanetary disks may be observed at centimeter wavelengths, and 
that changes in the spectral slope of the dust emission may be detected,
providing constraints on dust evolution and the planet formation process.

\end{abstract}


\section{Introduction}
Many young stars show thermal emission from solids in circumstellar 
disks with properties thought similar to the early Solar System.
Early analyses of spectral energy distributions indicated disk radii of 
10's to 100's of AU \citep{ada87} and masses of gas and dust sufficient 
to form planetary systems like our own \citep{bec90}. Statistical studies 
indicate disk dissipation on timescales of $\sim10$ Myr \citep{str89}, 
compatible with the standard paradigm of giant planet cores forming by 
dust coagulation followed by planetesimal accretion \citep{wuc00}.
A mysterious aspect of this process is that interstellar dust grains 
overcome energetic obstacles to sticking and grow to sizes large enough 
to decouple from the disk gas and interact gravitationally \citep{you02}. 
Most models postulate a phase of collisional agglomeration mediated by 
such mechanisms as brownian motion and turbulence. Numerical simulations 
of disk evolution indicate a bimodal distribution of particle sizes 
develops as millimeter size aggregates grow and settle to the 
disk mid-plane \citep[e.g.][]{wei97}, where the resulting dense layer 
becomes a reservoir for the formation of planetesimals and larger bodies.

Observations of dust emission over a wide range of wavelengths provide 
diagnostic information on particle properties \citep{bec91,bec00}. 
Recently recognized nearby stellar associations offer prime targets for 
observations to address the evolution of dust towards planets \citep{zuc01}.
The TW~Hya association, which contains more than 20 stellar systems with 
ages estimated to be 5 to 10 Myr, is the nearest known site of recent 
star formation activity \citep{son03}.
The classical T~Tauri star TW~Hya, an apparently single star of mass 
0.8~M$_{\odot}$, has 
become the focus of considerable attention because of its proximity 
(56 pc), and because it retains a remarkable face-on circumstellar 
disk of radius 225 AU visible in scattered light \citep{kri00,wei02} 
and in thermal emission from dust and trace molecules \citep{wei89,kas97,qi04}.

The dust in the TW~Hya disk, like the dust in disks around many younger 
stars, has long been known to show evidence for size evolution from a 
primordial interstellar distribution. At optical and near-infrared wavelengths,
the TW~Hya disk appears almost spectrally gray \citep{rob05},
indicative of grain growth 
from sub-micron sizes in the upper layers of the disk. At submillimeter 
and millimeter wavelengths, the spectral slope indicates a shallow dependence 
of the particle mass opacity on wavelength, $\kappa\sim\lambda^{\beta}$,
with $\beta\sim1$. For the TW~Hya disk, well resolved images of thermal 
dust emission confirm that optical depth effects do not affect the 
determination of the spectral slope \citep{wil00}.
For compact spherical particles of size $<< \lambda$, $\beta\rightarrow2$,
while for particles of size $>> \lambda$, $\beta\rightarrow0$. 
In realistic astrophysical mixtures, $\beta$ also partly depends on grain 
composition, structure, and topology, though the particle size generally 
dominates, especially for common silicates \citep{pol94,miy93}.
The shallow submillimeter spectral slope found for observations of TW~Hya 
has been robustly interpreted as evidence for particle growth to sizes of 
order 1 millimeter or more \citep{cal02,nat04}.

Since the particle size probed by observation is comparable to wavelength, 
detection of dust emission at centimeter wavelengths has the potential 
to reveal the development and location of larger ``pebbles'' within disks.
Two problems have historically thwarted this goal \citep{mun93}: 
(1) the opacity per unit mass decreases for larger particles and the 
resulting emission is weak, and 
(2) the centimeter wavelength emission of younger ($\sim1$ Myr old) stars 
is often dominated by hot plasma in the system, either gyrosynchrotron 
emission associated with chromospheric activity, or thermal bremsstralung 
emission from partially ionized winds. For TW~Hya, the close distance and 
the advanced age mitigate these problems.  
We present new observations of TW~Hya at centimeter wavelengths using 
the Very Large Array\footnote{The National Radio Astronomy Observatory 
is a facility of the National Science Foundation operated under cooperative 
agreement by Associated Universities, Inc.} (VLA) that convincingly argue 
for emission from a population of large dust particles in the disk.

\section{Observations and Results}
We used the VLA in late 2001 and early 2002 to observe TW~Hya at 
$\lambda$=3.5 cm at approximately bi-weekly intervals over 8 epochs 
in the D configuration (2001.732, 2001.770, 2001.808, 2001.847, 2001.885, 
2001.923, 2001.956, 2001.999), which provided $\sim8''$ (450~AU) maximum
resolution, and at 7 epochs in the A configuration
(2002.047, 2002.088, 2002.140, 2002.186, 2002.252, 2002.340, 2002.381),
which provided considerably higher $\sim0\farcs25$ (15 AU) resolution.
The typical rms noise in the images obtained from each of these short 
observations was $\sim40~\mu$Jy. In addition, we observed TW~Hya 
at $\lambda$=6~cm in the DnC configuration in 2001 September 24,
with beam size $\sim22''$  and rms noise 70 $\mu$Jy,
to provide supplemental spectral index information.
The AIPS software 
was used for calibration and imaging of each VLA data set following 
standard procedures. The systematic uncertainty in the absolute flux scales 
at these wavelengths is less than 10\%.

\subsection{3.5~cm emission not variable}
A previous VLA 3.5 cm $3\sigma$ upper limit of 84 $\mu$Jy \citep{ruc92} 
for TW~Hya compared to a subsequent detection at $200\pm28$~$\mu$Jy 
indicated that the emission at this wavelength was variable \citep{wil00}.
This apparent variability, together with the discovery that TW~Hya is a source 
of strong X-rays, suggested that the emission was due to hot gas resulting
from stellar magnetic activity, which is known to produce variations at 
radio wavelengths of an order of magnitude or more on timescales 
of hours to months to years \citep{kut86,whi92,chi96}.
Our initial impetus to monitor the TW~Hya 3.5~cm emission at frequent intervals
was to investigate the variability, and to determine if the radio flux
reached sufficient levels to be amenable to Very Long Baseline Interferometry.

\begin{figure}
\plotone{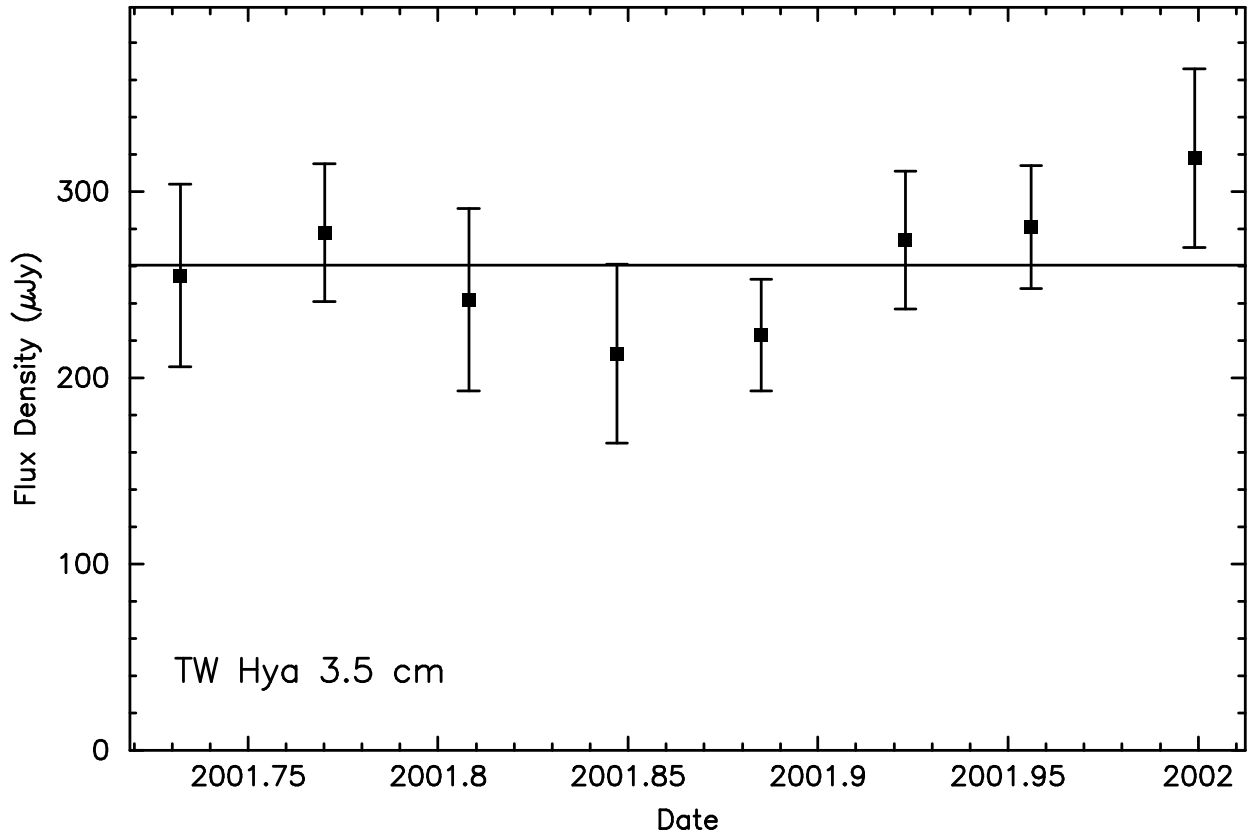}
\caption{The 3.5 cm continuum emission from TW~Hya shows no variability. 
The flux of the TW~Hya system was monitored at 8 epochs using the VLA in 
the D configuration with $\sim8''$ resolution, and the points show the 
measured fluxes with error bars that represent the rms noise in each 
observation. The solid line indicates the 260~$\mu$Jy mean value.
\label{fig:xband_time}
}
\end{figure}

Figure~\ref{fig:xband_time} shows the TW~Hya D configuration 3.5 cm 
measurements as a function of time. No variations from the 260 $\mu$Jy mean 
were observed within the $\sim40~\mu$Jy rms uncertainties at each epoch. 
By contrast, a second radio source within the field of view with flux 
comparable to TW~Hya, presumably a background active galactic nucleus, showed 
variations by more than a factor of two during this period. 

The surprising lack of variability for the TW~Hya radio emission prompted us 
to check the previously reported upper limit. A re-analysis of these data from 
1991 January 30 in the VLA archive shows that the noise was significantly 
higher than reported, and excision of bad data results in a detection of 
TW~Hya at the $210\pm55~\mu$Jy level. The early observations that 
were thought to indicate variability in fact are compatible with all of 
the subsequent (constant) measurements.

\subsection{3.5 cm emission spatially resolved}
\begin{figure}
\plotone{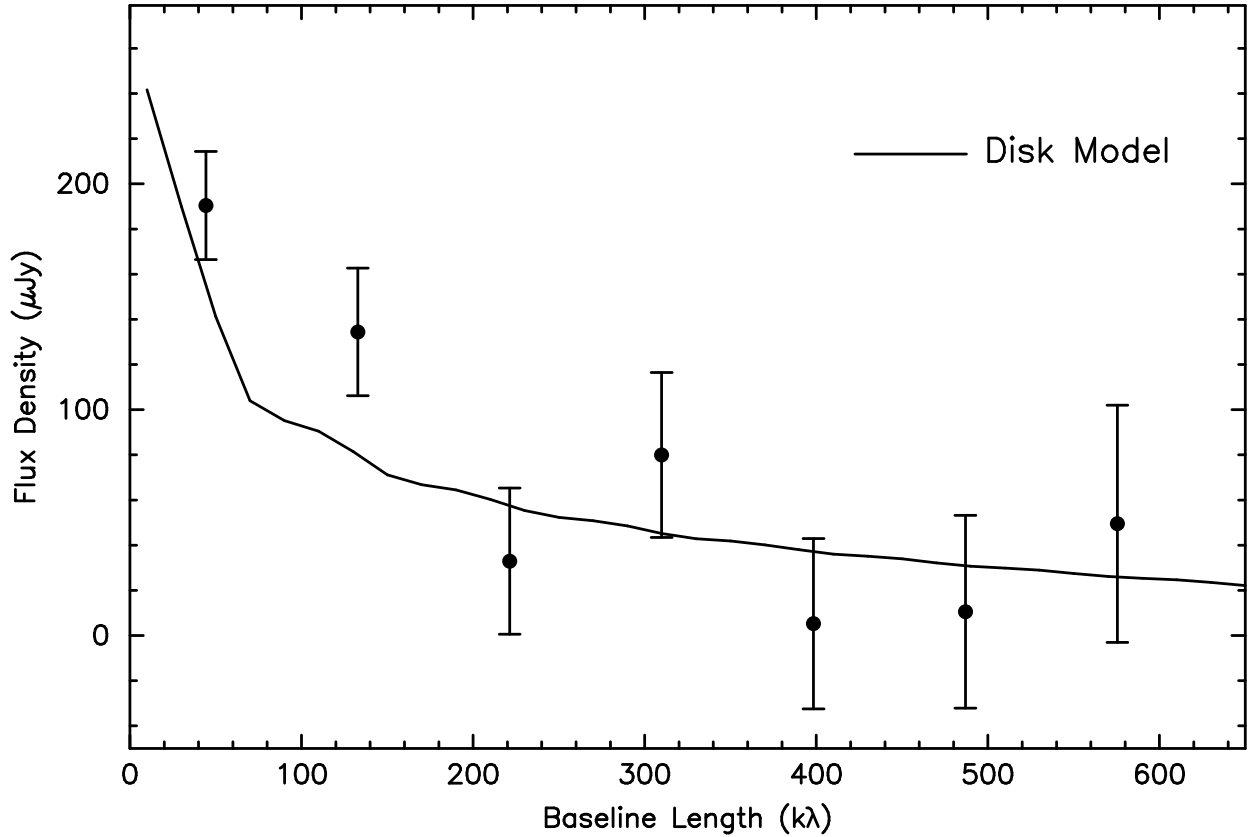}
\caption{The 3.5 cm continuum emission from TW~Hya is spatially resolved  
by observations with a subarcsecond beam. The plot is derived from the 
sum of observations obtained at 7 epochs with the VLA in the A configuration. 
The points show the real part of the 3.5 cm visibility as a function of 
baseline length, annularly averaged in bins of width 88 k$\lambda$. 
The error bars represent $\pm1$ standard deviation for each bin. 
The solid curve shows the predicted emission from an irradiated accretion 
disk with a population of large dust particles (see \S~\ref{sec:diskmodel}).
\label{fig:xband_resolved}
}
\end{figure}

The A configuration 3.5 cm TW~Hya observations spatially resolve the 
emission region. Because the brightness distribution is strongly centrally 
peaked and falls off steeply with radius, this is best illustrated by 
examination of the interferometer data in the visibility domain. 
Figure~\ref{fig:xband_resolved} shows the real part of the 3.5 cm visibility 
as a function of baseline length, annularly averaged in bins of 88 k$\lambda$ 
width, combining data from all of the observed epochs to maximize the
signal-to-noise ratio.  The falloff in visibility at longer baselines is 
the signature that the emission is spatially resolved.

\section{Discussion}

\subsection{Dust Emission at 3.5 cm}
The lack of time variability, together with spatial extent orders of 
magnitude larger than a stellar radius, provide strong constraints on the 
emission mechanism. In particular, chromospheric activity-- expected to arise 
from transient magnetic features comparable in extent to the star-- 
is not responsible for the 3.5~cm emission. More detailed X-ray data bolster 
this conclusion. Observations of the X-ray spectrum with 
XMM-Newton \citep{ste04} and Chandra \citep{kas02} indicate electron 
densities two orders of magnitude higher than typically found in stellar 
coronae. In addition, the X-ray emitting material appears to be depleted of 
grain-forming elements like Mg and Si. The X-ray emission is more likely the
result of accretion shocks at the star/disk interface. Any radio emission 
from such accretion shocks would be on such small size scales that spatial
extent would not be resolved by the VLA observations.

An ionized wind does not present a viable explanation for the 3.5 cm emission,
either. While there is evidence for small amounts of ionized gas in the 
TW~Hya system provided by optical spectroscopy of e.g. H$\alpha$ emission 
\citep{muz00}, there are no indications from direct imaging of a wind in the 
form of an optical jet, as found in younger stars with higher disk accretion 
rates. A persuasive scaling argument against significant wind emission 
has been made previously from TW~Hya's low inferred accretion rate of
$\sim10^{-9}$~M$_{\odot}$~yr$^{-1}$, which suggests a low mass outflow rate, 
and therefore a 3.5 cm flux more than an order of magnitude lower than 
observed \citep{cal02}.
The new VLA observations provide two additional arguments against a wind 
origin for the emission.  First, the $3\sigma$ upper limit at 6~cm 
constrains the spectral spectral index from 6 to 3.5 cm to be $>0.3$, which 
is incompatible with the nearly flat value of optically thin ionized gas, and 
steeper than typically found for the winds from young stars \citep{ang98}. 
Second, and more important, the peak brightness temperature at 3.5 cm in the 
resolved data is only $\sim10$~K, which is compatible with ionized gas 
only if the optical depth or filling factor of optically thick emission 
is very low, unlike any known ionized jet from a pre-main-sequence star. 

Dust emission naturally accounts for all of the properties 
of the 3.5 cm emission: (1) lack of time variability,
(2) large spatial extent,
(3) low brightness at high spatial resolution, 
and (4) the spectral index constraint.
We conclude that the 3.5~cm emission from TW~Hya is strongly dominated 
by thermal emission from dust particles in the disk surrounding the star.

\subsection{A Disk Model}
\label{sec:diskmodel}
\begin{figure}
\plotone{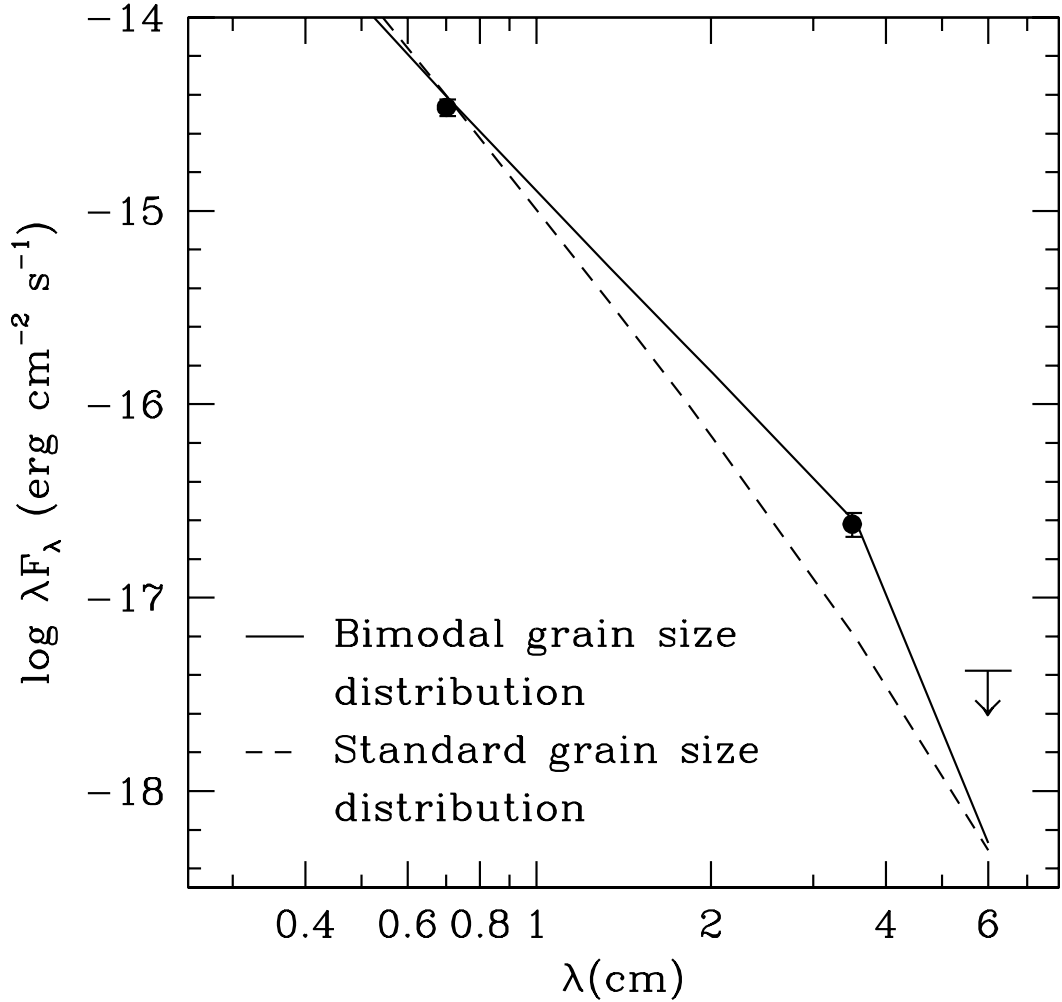}
\caption{The filled circles show VLA measurements at 7 mm and 3.5 cm, 
and the arrow represents an upper limit at 6 cm.
The long wavelength spectrum of TW~Hya is better fit by a model 
in which nearly all of the mass in solids is in centimeter size 
grains (solid line) than by the model of \citet{cal02} based on 
a single-power power law grain size distribution (dashed line).  
\label{fig:twhya_spectrum}
}
\end{figure}

The full spectrum of TW~Hya from the infrared through the millimeter, 
has been matched very well by a self-consistent irradiated accretion disk 
model that assumes standard optical properties for a mix of spherical 
particle constituents and a single-power power law size distribution 
\citep{cal02}.   This model does not fit the observations at 3.5 cm, however. 
The dashed curve in Figure~\ref{fig:twhya_spectrum} shows that 
this previous model falls short of the observations at 3.5 cm by a factor 
of a few, which is much larger than the observational uncertainty.
The basic reason is that many more grains that emit efficiently at centimeter 
wavelengths must be included in order to raise the level of dust emission 
at long wavelengths.

A disk model that includes a plausible population of large grains can match 
the data. For illustration, we assume a simple model 
that contains two grain populations: (1) small grains with a size distribution 
(as a function of particle radius $a$) 
$n(a) \sim a^{-3.5}$ between minimum and maximum sizes 
$a_{min}=0.005$ $\mu$m and $a_{max}=1$ $\mu$m, and (2) larger ``pebbles''
with size distribution $n(a) \sim a^{-2.5}$, between minimum and maximum
sizes $a_{min}=5$ mm and $a_{max}=7$ mm. The small grains, which contain 
only 0.1\% of the particle mass in the disk, are needed to regulate the
irradiated disk structure and to explain the infrared spectrum, including 
the silicate feature near 10~$\mu$m \citep{dal01}. 
The solid curve in Figure~\ref{fig:twhya_spectrum} shows that the model 
with large grains boosts the long wavelength emission to match 
the observed flux density. 
The same model also matches very well the observed brightness distribution.
Figure~\ref{fig:xband_resolved} shows the Fourier transform of the 
brightness distribution derived from this model as a solid curve, and
the model follows the falloff in the visibilities. 
Note that the large grains that account for the 3.5 cm emission must be 
present to radii of at least 10's of AU to match the observed brightness 
distribution. The disk mass in this model is 
0.1~M$_{\odot}$, assuming a standard dust-to-gas mass ratio. 
This disk mass lies below the limit beyond which gravitational 
instabilities are expected to set in.

The parameters in this simple model are clearly not unique. 
The strictly bimodal and discontinuous particle size distribution used 
in this disk model is meant to be illustrative of grain growth and settling,
and to capture the essential behavior of more sophisticated treatments
that result in small grains floating above a dense layer in the mid-plane
\citep[e.g.][]{wei97, tan05}. 
The detailed situation is complex, with faster particle growth in the 
inner disk, and competition between growth, settling and turbulence, 
at all disk radii. 
The new observations, which show elevated emission from dust at 3.5~cm 
and a {\em change} in the spectral slope of the dust emission from
submillimeter to centimeter wavelengths, are difficult to explain unless 
substantial particle evolution has occurred throughout the disk. 
Future observations with higher signal-to-noise at high resolution 
hold the promise to map the spectral index of the dust emission in the 
disk and locate regions of changing grain properties. 

\subsection{Planet Formation}
Independent of the details of the disk model, the detection of dust emission 
at centimeter wavelengths provides a strong indications that an early phase 
of the planet building process is underway in the TW~Hya disk.
The inference of a $\sim4$~AU radius inner hole in dust disk from the
mid-infrared spectral energy distribution has previously given rise to the 
suggestion that a planet may be present in the TW~Hya system 
\citep{cal02}.
In this context, it is interesting to consider that changes in dust properties 
may affect the timescale for planet formation. A key result in recent studies 
of giant planet formation by core accretion is the extraordinary sensitivity 
of the time of onset of a rapid gas accretion phase to assumptions of 
disk surface density and dust opacities; for example, \citet{hub04} show 
that decreasing dust opacity can reduce the formation time of gas giant 
planets by more than a factor of two from the canonical 8~Myr of older
models \citep{pol96}.
This shorter giant planet formation timescale is more comfortably consistent 
with estimates for the age of the TW~Hya system.  

\subsection{Radio Emission from TW Hya in Context}
It would be extremely interesting if other classical T Tauri stars 
showed invariate centimeter emission with spectral 
properties indicative of dust emission like TW Hya. Unfortunately, existing
VLA surveys of the large sample of sources associated with the most
nearby dark clouds detect only younger Class I objects or transition 
objects with high accretion rates that drive radio and optical jets, 
like HL Tau and DG Tau \citep{coh82,bie84}, 
and weak-lined T Tauri stars with stellar activity \citep[e.g.][]{chi96}. 
For classical T Tauri stars, the upper limits at 6 cm are generally 
160-300 $\mu$Jy.  A few sources have comparable or better limits at 3.5~cm, 
e.g. DO Tau $<170$ $\mu$Jy \citep{koe95} and DL Tau and GG Tau $<77$ $\mu$Jy 
\citep{mun93}.
If TW Hya were at 140~pc, then it would be among the stronger dust disks
at millimeter wavelengths, but its 3.5 cm emission of only $\sim40$~$\mu$Jy
would lie well below the existing detection limits for sources at that distance.
More sensitive observations at a range of centimeter wavelengths are needed
to determine what fraction, if any, of the disks around classical T Tauri 
stars show evidence for extreme grain growth like that inferred for TW Hya.

The detection of dust emission at centimeter wavelengths from the TW~Hya 
disk provides direct evidence for much larger particles than have been 
detected before in a nebula like the one from which the Solar System is 
thought to have emerged. This first observation points the way to a new 
field of investigation.  Next generation radio telescopes, in particular 
the the Square Kilometer Array \citep{laz04}, will have a combination of 
sensitivity and angular resolution sufficient to image in detail the 
distribution of large dust particles emitting at centimeter wavelengths in 
the disks around hundreds of young stars in nearby dark clouds and beyond. 

\acknowledgments
Partial support for this work was provided by NASA Origins of Solar Systems 
Program Grants NAG5-11777 and NAG5-9670.

\clearpage

\end{document}